\documentclass[lettersize,journal]{IEEEtran}
\usepackage{amsmath,amsfonts}
\usepackage{amssymb}
\usepackage[caption=false,font=footnotesize,labelfont=sf,textfont=sf]{subfig}
\usepackage{stfloats}
\usepackage{url}
\usepackage{graphicx}
\usepackage{cite}
\usepackage{float}
\usepackage{subfig}
\usepackage{multirow}
\usepackage{color}
\usepackage{makecell}
\usepackage{booktabs}
\usepackage{mathtools}
\usepackage[colorlinks,citecolor=blue,linkcolor=blue]{hyperref}

\newtheorem{theorem}{Theorem}

\newtheorem{lemma}{Lemma}

\usepackage{algorithm, algorithmicx, algpseudocode}
\begin{document}

\title{Capacity Bounds and Low-Complexity Constellation Shaping under Mixed Gaussian-Impulsive Noise}

%%%%%%
\author{Tianfu Qi, \emph{Graduate Student Member, IEEE}, Jun Wang, \emph{Senior Member, IEEE}
\thanks{Tianfu Qi, Jun Wang are with the National Key Laboratory on Wireless Communications, University of Electronic Science and Technology of China, Chengdu 611731, China (e-mail: 202311220634@std.uestc.edu.cn).}
}

% The paper headers
\markboth{Journal of \LaTeX\ Class Files,~Vol.~14, No.~8, August~2021}%
{Shell \MakeLowercase{\textit{et al.}}: A Sample Article Using IEEEtran.cls for IEEE Journals}

\maketitle

\begin{abstract}
This paper investigates the bounds on channel capacity and constellation shaping under memoryless mixed noise, which is composed of impulsive noise (IN) and white Gaussian noise (WGN). The capacity bounds are derived using the entropy power inequality and the dual expression of capacity. It is then shown that the proposed lower and upper bounds asymptotically converge to the true channel capacity, and the analytic asymptotic capacity expression is obtained. Leveraging this property, we design a low-complexity constellation shaping method that operates without iterative procedures. Simulation results demonstrate that the derived bounds are remarkably tight, and the shaped constellation achieves the highest mutual information among all considered baseline schemes.
\end{abstract}


\vspace{-0.3cm}
\section{Introduction}
Impulsive noise (IN) is present in a variety of real-world communication systems, such as underwater acoustic communications \cite{paper1,paper2}, IoT wireless networks \cite{paper3}, NOMA systems \cite{paper4}, and smart grids \cite{paper5}. A key characteristic of IN in the time domain is its large amplitude and high power over short durations. In practice, such noise often occurs alongside ever-present white Gaussian noise (WGN), forming a mixed noise environment. If IN is ignored and conventional algorithms designed solely for WGN are applied directly, communication systems can experience significant performance degradation \cite{paper9,paper38,paper39}.

An effective approach to enhance system performance is through shaping the transmitted constellation points. For the WGN channel, it is well established that conventional square constellations (e.g., M-ary QAM signals) are suboptimal, with the maximum achievable shaping gain reaching 1.53 dB for sufficiently high modulation orders \cite{paper24}. Intuitively, constellation shaping aims to maximize the minimum Euclidean distance between all constellation point pairs. In the two-dimensional (2D) case, this leads to a hexagonal distribution, which is also known as the optimal solution to 2D packing and covering problems \cite{paper11}. However, in more complex communication systems and channels, constellation shaping is often formulated as an optimization problem and solved using iterative numerical algorithms. Constellation optimization under WGN has been studied in various scenarios, including underwater wireless optical communication \cite{paper19}, ambient IoT networks \cite{paper20}, and ISAC systems \cite{paper21}.

To the best of our knowledge, few studies have addressed constellation shaping under mixed channel noise conditions. While this problem can generally be formulated as maximizing certain information measures or minimizing detection error probability, such approaches often entail high computational complexity, limiting their practical applicability. Inspired by \cite{paper24}, geometric shaping (GS) could be directly implemented if the optimal input distributions for specific channels were known. Unfortunately, this is generally infeasible, as both the channel capacities and the corresponding capacity-achieving input distributions remain unsolved problems in such scenarios. Therefore, a more practical and tractable approach is to analyze capacity bounds and their asymptotic behaviors. The ideal case would be for the lower and upper bounds to converge across all signal-to-noise ratio (SNR) values.

Motivated by these considerations, the main contributions of this work are twofold and summarized as follows:
\begin{itemize}
\item{First, we derive tight lower and upper bounds for the channel capacity under mixed noise conditions. The lower bound is obtained directly using the entropy power inequality. Subsequently, we establish the upper bound via the duality expression of capacity with an appropriately chosen output distribution. More importantly, we demonstrate that the gap between the lower and upper bounds diminishes as the SNR approaches infinity. In this asymptotic regime, we obtain the closed-form capacity expression.}
\item{Second, leveraging the obtained input distribution, we implement GS without iterative procedures. Additionally, probabilistic shaping (PS) is incorporated to further enhance the mutual information (MI) of the optimized transmit signals. Moreover, both the GS and PS approaches feature analytical implementation procedures, ensuring low computational complexity and thus offering high practicality. Simulation results confirm that the proposed shaping scheme achieves higher MI compared to existing methods.}
\end{itemize}

\vspace{-0.3cm}
\section{Preliminary}
In this section, we first outline the channel noise model adopted in this work, then describe the considered constraints and formulate the capacity problem.

\vspace{-0.2cm}
\subsection{Channel Noise Model}
Memoryless mixed channel noise is commonly modeled as the sum of WGN and memoryless IN. Since they typically originate from independent sources, it is reasonable to assume they are mutually independent. Accordingly, the mixed noise model is expressed as follows,
\begin{align}\label{mixed_noise_model}
N=N_G+N_I,
\end{align}
where $N$, $N_G$, and $N_I$ denote the random variables (RVs) of the mixed noise, WGN and IN, respectively. The noise is assumed to have zero mean, and thus the amplitude of $N_I$ follows a symmetric $\alpha$-stable (S$\alpha$S) distribution \cite{paper38,paper39}. Unfortunately, in this case, $N$ does not possess a closed-form probability density function (PDF), posing significant challenges for practical applications. To address this issue, our prior work \cite{paper9} introduced an approximate PDF for $N$, given as follows,
\begin{align}\label{GS_model}
f_{N}(n)=\rho k_1\exp\left(-\frac{n^2}{4\gamma_g^2}\right) +\frac{(1-\rho)k_2}{(1+n^2/(2\alpha\gamma_s^2))^\frac{\alpha+1}{2}},
\end{align}
where $k_1$ and $k_2$ are normalization factors, given by $k_1=(2\sqrt{\pi}\gamma_g)^{-1}$, $k_2=\Gamma((\alpha+1)/2)/(\Gamma(\alpha/2)\sqrt{2\alpha\pi\gamma_s^2})$ where $\Gamma(\cdot)$ is the Gamma function. The model in \eqref{GS_model} contains four parameters. The characteristic parameter $\alpha \in (0,2]$ controls the heaviness of the PDF's tail. The scaling parameters $\gamma_s \in [0, +\infty)$ and $\gamma_g \in [0, +\infty)$ correspond to the IN and WGN components, respectively. The weight parameter $\rho \in [0,1]$ determines the relative strength of WGN and IN in the mixture. It has been shown that \eqref{GS_model} is able to accurately describe the real-world noise samples. Further details regarding the model's derivation and validation are available in \cite{paper9}.

\vspace{-0.3cm}
\subsection{Problem formulation}
Consider the following received signal model:
\begin{align}\label{received_signal_model}
Y=X+N,
\end{align}
where $X$, $Y$ and $N$ represent the RVs of the transmitted signal, received signal and channel noise, respectively. Let $F_X(\cdot)$ denote the cumulative distribution function (CDF) of $X$, subject to the power constraint:
\begin{equation}\label{constraint_1}
\mathbb{E}_{X}[|X|^p]\leq P_0<+\infty,1\leq p<\alpha\leq2,
\end{equation}
where $\mathbb{E}_{X}[\cdot]$ denotes the expectation with respect to $X$. The lower bound on $p$ in \eqref{constraint_1} ensures the convexity of the constraint. We employ the $p$-th norm rather than variance as it constitutes a more general and weaker power measure. Generally, a finite variance implies the finite $p$-th moment, but the converse does not hold. Moreover, the $p$-th norm ($0<p<+\infty$) equivalently characterizes the signal power from a functional perspective. It follows that the output distribution $F_Y(y)$ must also be power-limited. The feasible set $\Omega$ of input distributions is thus defined as $\Omega=\{F_X:\mathbb{E}_{X}[|x|^p]\leq P_0,\forall x\in \mathbb{R},1\leq p <\alpha\leq2\}$.

Finally, we state a simple lemma regarding the $p$-th moment of the output without proof.
\begin{lemma}\label{lemma_1}
\textit{Assume the input $X$ satisfies the power constraint \eqref{constraint_1}, and the amplitude of the channel noise follows \eqref{mixed_noise_model}. Then, $\mathbb{E}_{Y}[|Y|^p]^{\frac{1}{p}}\leq P_0^{\frac{1}{p}}+\mathbb{E}_{N}[|N|^p]^{\frac{1}{p}}<+\infty$ for arbitrary $p<\alpha$.}
\end{lemma}

\section{Capacity Bounds}
For the mixed noise channel, obtaining an explicit channel capacity expression is highly challenging due to the complex PDF form, the absence of finite second-order moments, and other complications. In this section, we first derive lower and upper bounds on the capacity under the given constraints. We then analyze the asymptotic convergence of these bounds, which facilitates their practical application.

\vspace{-0.2cm}
\subsection{Lower bound}
The lower bound is relatively straightforward to derive using the entropy power inequality (EPI), i.e.,
\begin{equation}\label{entropy_power_inequality}
C\geq \frac{1}{2}\log\left(1+e^{2[h(X)-h(N)]}\right)\triangleq C_L.
\end{equation}

Thus, a tighter lower bound can be obtained by maximizing $h(X)$ under constraints \eqref{constraint_1}. Following the maximum entropy principle, the optimal distribution $f_X(x)$ follows the form:
\begin{equation}\label{optimal_input}
f_X(x)=e^{\lambda_0+\lambda_1|x|^p},\forall x\in \mathbb{R}.
\end{equation}

Combining this result with constraint \eqref{constraint_1}, we can obtain that $\lambda_0=(p-1)\log p/p-p^{-1}\log P_0-\log\left[2\Gamma\left(1/p\right)\right]$, $\lambda_1=-(pP_0)^{-1}$ and $h(X)=-\lambda_0+p^{-1}$. Therefore, the channel capacity lower bound is given by:
\begin{align}
C_L=\frac{1}{2}\log\left(1+e^{2[-\lambda_0+1/p-h(N)]}\right).
\end{align}

As shown in the following subsection, equality in the EPI is achievable, and this lower bound converges to the channel capacity with increasing signal power.

\vspace{-0.3cm}
\subsection{Upper bound}
In \cite{paper14}, a discrete duality inequality based on relative entropy was established and later extended to a continuous form equivalent to an infinite input alphabet. The resulting generalized upper bound is given by \cite{paper14}:
\begin{equation}\label{relative_inequality}
C\leq \mathbb{E}_{X}\left[\mathcal{K}\left(f_{N}(y|x);f_R(y)\right)\right],
\end{equation}
where $f_R(y)$ is an arbitrary output distribution to generate the upper bound. The equality is achieved when $f_R(y)=f_Y(y)$ and $f_Y(y)$ is the output PDF corresponding to the optimal input. $\mathcal{K}(\cdot;\cdot)$ denotes the KL divergence between two distributions. An elaborated $f_R(y)$ will lead to a tight upper bound with a closed form.

The channel capacity is formulated as an optimization problem to determine the optimal input distribution. Since the channel noise entropy is independent of the input, the problem can be reformulated as
\begin{align}\label{duality_optimized_problem}
C=&\sup\limits_{F_X \in \Omega}\left\{h(Y)-h(Y|X)\right\}=\sup\limits_{F_X \in \Omega}\left\{h(Y)\right\}-h(N)\nonumber\\
=&\sup\limits_{F_Y \in \tilde{\Omega}}\left\{h(Y)\right\}-h(N).
\end{align}
where $\tilde{\Omega}$ represents the feasible set of $Y$. The optimization problem in \eqref{duality_optimized_problem} can be interpreted as a dual formulation aimed at identifying the optimal output distribution under the given constraints. Considering \eqref{constraint_1}, a power constraint also applies to $Y$. However, except for the Gaussian case ($p=2$), the exact power constraint for $Y$ cannot be explicitly determined.

According to Lemma \ref{lemma_1}, the $p$-th moment of $Y$ remains finite. Therefore, we have $\tilde{\Omega}_Y\triangleq\{F_Y:\mathbb{E}_{Y}[|Y|^p]\leq P_Y,\forall y\in \mathbb{R},1\leq p <\alpha\leq2\}\subset\tilde{\Omega}$ where $P_Y$ is given in Lemma \ref{lemma_1}. This allows us to search for a solution to problem \eqref{duality_optimized_problem} within the more tractable set $\tilde{\Omega}_Y$. It will be shown that this relaxation of the feasible set is sufficient to obtain an asymptotically tight upper bound. The above analysis motivates the choice of an appropriate distribution for $f_R(\cdot)$ in \eqref{relative_inequality}, leading to the following theorem.

\begin{theorem}\label{theorem_1}
\textit{Let the arbitrary output distribution $f_R(y)$ be
\begin{equation}\label{upper_1_distribution}
f_R(y)=\frac{p}{2\Gamma\left(1/p\right)}\sigma_1^{-\frac{1}{p}}e^{-\frac{|y|^p}{\sigma_1}},\forall y\in \mathbb{R},\sigma_1>0,
\end{equation}
where $\sigma_1$ is the scaling parameter of $f_R(y)$. Then, the capacity is upper-bounded as follows,
\begin{align}\label{upper_bound_1}
C\leq& -\log\frac{p}{2\Gamma\left(1/p\right)}+\log\Big(P_0^{\frac{1}{p}}+\mathbb{E}_{N}\left[|N|^p\right]^{\frac{1}{p}}\Big)\nonumber\\
&+\frac{1}{p}+\frac{1}{p}\log p-h(N)\triangleq C_U.
\end{align}}
\end{theorem}

\begin{IEEEproof}
Based on \eqref{relative_inequality}, we can obtain that
\begin{align}
C\leq&\inf\limits_{\sigma_1}\left\{-\mathbb{E}_{X}\left[\int_{-\infty}^{+\infty}f_{N}(y|x)\log f_R(y)dy \right]-h(N)\right\}. \label{upper_1_1}
\end{align}

Substituting \eqref{upper_1_distribution} into \eqref{upper_1_1} gives
\begin{align}
&-\mathbb{E}_{X}\left[\int_{-\infty}^{+\infty}f_{N}(y|x)\log f_R(y)dy \right] \nonumber\\
=&-\log\Big(\frac{p}{2\Gamma(1/p)}\sigma_1^{-\frac{1}{p}}\Big)+\frac{1}{\sigma_1}\mathbb{E}_{X}\Big[\int_{-\infty}^{+\infty}f_{N}(n)|x+n|^pdn\Big].
\end{align}

Then, according to Minkovski's inequality,
\begin{align}\label{upper_1_derive_1}
&\mathbb{E}_{X}\Big[\int_{-\infty}^{+\infty}f_{N}(n)|x+n|^pdn\Big]\nonumber\\
\leq&\mathbb{E}_{X}\Big[\Big(\mathbb{E}_{N}\left[|X|^p\right]^{\frac{1}{p}}+\mathbb{E}_{N}\left[|N|^p\right]^{\frac{1}{p}}\Big)^p\Big]\nonumber\\
\leq&\Big(P_0^{\frac{1}{p}}+\mathbb{E}_{N}\left[|N|^p\right]^{\frac{1}{p}}\Big)^p\triangleq\hat{U}_0.
\end{align}

Consequently, \eqref{upper_1_1} can be further written as
\begin{equation}
C\leq \inf\limits_{\sigma_1}\bigg\{\underbrace{-\log\frac{p}{2\Gamma\left(1/p\right)}+\frac{1}{p}\log\sigma_1+\frac{\hat{U}_0}{\sigma_1}}_{\hat{U}_1(\sigma_1)}-h(N)\bigg\}.
\end{equation}

The optimal value of $\sigma_1$ can be determined by taking the first derivative of $\hat{U}_1(\sigma_1)$. Note that $\hat{U}_0$ is independent of $\sigma_1$, and then, we have $\sigma_1^*=p\hat{U}_0$. After substitution and simplification, the first capacity upper bound is obtained as shown in \eqref{upper_bound_1}.
\end{IEEEproof}

\vspace{-0.2cm}
\subsection{Asymptotic analysis}
Next, we analyze the asymptotic relationship between $C_L$ and $C_U$, noting that the input distribution for $C_L$ and the output distribution for $C_U$ are identical. Denoting the difference as $\Delta B\triangleq C_U-C_L$. As the $P_0$ continuously increases such that $P_0\gg\mathbb{E}_{N}\left[|N|^p\right]$, we have
\begin{align}
\Delta B=&-\log\frac{p}{2\Gamma\left(1/p\right)}+\frac{1}{p}\log P_0+\frac{1}{p}\log p+\frac{1}{p}\nonumber\\
&-\frac{1}{2}\log\big(1+e^{2[1/p-\lambda_0-h(N)]}\big).
\end{align}

According to the previous analysis, $-\lambda_0$ is also large enough. Consequently,
\begin{align}
\Delta B\rightarrow& -\log\frac{p}{2\Gamma\left(1/p\right)}+\frac{1}{p}\log P_0+\frac{1}{p}\log p+\frac{1}{p}\nonumber\\
&-\Big(\frac{1}{p}-\lambda_0-h(N)\Big)\triangleq\Delta B(P_0).
\end{align}

Therefore, $\lim\limits_{P_0 \to +\infty} \Delta B(P_0) = 0$, indicating that $C_L$ and $C_U$ converge to the same value at high input power. Furthermore, the asymptotic channel capacity under constraints \eqref{constraint_1} is given by:
\begin{align}\label{analytic_channel_capacity_expression}
\lim\limits_{P_0\rightarrow+\infty} C(p,P_0)=-\log\frac{p}{2\Gamma\left(1/p\right)}+\frac{\log(pP_0)+1}{p}-h(N),
\end{align}
where $1\leq p<\alpha\leq2$. We now consider the case where $\alpha \rightarrow 2$ to verify the correctness and generality of \eqref{analytic_channel_capacity_expression}. Indeed, when the input signal power is sufficiently large,
\begin{align}
\lim\limits_{\alpha\rightarrow2} C(2,P_0)=&\log\Gamma\left(1/2\right)+\frac{\log(2eP_0)}{2}-\lim\limits_{\alpha\rightarrow2}h(N)\nonumber\\
=&\frac{1}{2}\log\left(2\pi eP_0\right)-\frac{1}{2}\log(4\pi e\gamma_{g}).
\end{align}

Based on \eqref{mixed_noise_model}, $2\gamma_{g}$ is equivalent to the noise variance with $\alpha=2$ and $c_1=1$. Thus, we have $\lim\limits_{\alpha\rightarrow2,P_0\rightarrow+\infty} C(2,P_0)=\lim\limits_{P_0\rightarrow+\infty}\frac{1}{2}\log(P_0/P_N)$ where $P_N$ denotes the noise power. It coincides with the Shannon formula as $P_0 \rightarrow +\infty$, confirming the generality of \eqref{analytic_channel_capacity_expression}.

$\textbf{Remarks}$: According to Theorem \ref{theorem_1}, the parameter $p$ also affects the tightness of the capacity bounds. However, the expression for the optimal $p$ is complex and requires time-consuming numerical optimization. Therefore, we do not optimize over $p$, and the influence of different $p$ values on the bounds is beyond the scope of this paper.

\vspace{-0.2cm}
\section{Low-complexity constellation shaping}
Conventional joint geometric and probabilistic shaping is typically achieved through iterative optimization algorithms, which entail high computational complexity. Leveraging the derived asymptotic channel capacity and its corresponding optimal input distribution, we propose a low-complexity shaping algorithm that operates without iteration.

\vspace{-0.3cm}
\subsection{Geometric shaping}
For the WGN channel, geometric constellation shaping is typically implemented by quantizing a Gaussian distribution according to the Lloyd criteria. This approach is proven to be capacity-achieving at sufficiently high modulation orders. The key to this method lies in obtaining the optimal input distribution, which is straightforward for WGN but often intractable for other channels.

With the aid of \eqref{optimal_input} and Theorem \ref{theorem_1}, we can now design a corresponding geometric shaping scheme for the mixed noise channel. Although the capacity bounds were derived for the one-dimensional case, they can be directly extended to multidimensional scenarios. Accordingly, we adopt the input distribution given in \eqref{upper_1_distribution}. While this distribution is not strictly optimal in all regimes, it is asymptotically capacity-achieving under the mixed noise channel.

As for the quantization procedure, we omit the detailed description due to space constraints. A straightforward approach is to independently quantize each axis using the one-dimensional method and combine the results via the Kronecker product. For further quantization methods and related analyses, we refer the reader to \cite{book1,paper12}.

\vspace{-0.3cm}
\subsection{Probabilistic shaping}
Following reasoning similar to \cite{paper24}, the quantized discrete distribution described in the previous subsection can be shown to approach the continuous original. Consequently, the corresponding mutual information asymptotically converges to that of the continuous input distribution.

However, in practical systems with finite modulation orders, a gap remains between the channel capacity and the mutual information achieved by the shaped constellation. Although this finite-order quantization breaks strict asymptotic optimality, probabilistic shaping remains a viable method for further improving mutual information in real-world scenarios. The corresponding optimization problem is formulated as follows:
\begin{align}\label{PS_optimization}
(p_1^M)^*=\arg\max\limits_{p_1^M\in\Omega_p}I(Y;X)=\arg\max\limits_{p_1^M\in\Omega_p}h(Y)
\end{align}
where $\Omega_p=\{p_1^M:\sum_{j=1}^{M}p_j\Vert \mathbf{x}_j\Vert^p\leq P_0\}$ is the feasible set of the input distribution and $p_1^M\triangleq(p_1,\cdots,p_M)$. Here, $\mathbf{x}_j$ represents the $j$-th complex-valued constellation point with probability $p_j$. Slightly abusing notation, we denote the input entropy by $h(X)$, even though $X$ follows a discrete distribution. Due to the coupling of the $p_j$ in the logarithm, problem \eqref{PS_optimization} generally requires numerical methods such as the Blahut-Arimoto (BA) algorithm \cite{paper40}.

To develop a PS scheme with analytical expressions, we first present several observations. First, PS becomes unnecessary in the high-SNR regime. In this case, $I(Y;X) = h(X) - h(X|Y) \approx h(X)$ due to the negligible channel noise. Thus, maximizing $I(Y;X)$ is approximately equivalent to maximizing $h(X)$, which is achieved by a uniform input distribution. In the low-SNR regime, $h(X|Y)$ remains nearly constant, as it is dominated by the noise entropy, and little information about $X$ can be extracted from $Y$. However, $\max_{p_1^M \in \Omega_p} I(Y;X)$ is not equivalent to simply maximizing $h(X)$ here. Following the approach for the WGN channel, we may ignore the variation in $h(X|Y)$ and solve the power-constrained entropy maximization problem. Applying the Lagrange method with constraint \eqref{constraint_1}, we obtain:
\begin{align}\label{optimal_probility_distribution}
p_j^*\propto\exp(\Vert \mathbf{x}_j\Vert^p/s)
\end{align}
where $s$ is a scaling hyperparameter. \eqref{optimal_probility_distribution} clearly reduces to the Maxwell-Boltzmann (MB) distribution when $p=2$ \cite{paper15}. However, PS operation generally reduces the total constellation power. If power scaling is applied, the resulting distribution no longer matches the geometric positions. To address this issue, we introduce an additional projection step. Specifically, let $\tilde{p}_j=\exp(\Vert \mathbf{x}_j\Vert^p/\delta)/\sum_{j=1}^{M}\exp(\Vert \mathbf{x}_j\Vert^p/\delta)$. We then project $\tilde{p}_1^M$ onto the feasible set $\Omega^*_p=\{p_1^M:\sum_{j=1}^{M}p_j\Vert \mathbf{x}_j\Vert^p=P_0,\sum_{j=1}^{M}p_j=1\}$. After introducing the Lagrange multiplier, we should solve
\begin{align}\label{problem}
(p_1^M)^*=\arg\max\limits_{p_1^M\in\Omega^*_p}L(p_1^M,\tilde{p}_1^M)
\end{align}
where $L(p_1^M,\tilde{p}_1^M)=\frac{1}{2}\sum_{j=1}^{M}|p_j-\tilde{p}_j|^2+\lambda(\sum_{j=1}^{M}p_j\Vert \mathbf{x}_j\Vert^p-P_0)+\delta(\sum_{j=1}^{M}p_j-1)-\sum_{j=1}^{M}\beta_jp_j$ and $\lambda$, $\delta$, $\beta_j,j=1,\cdots,M$ are multipliers. An analytical solution to problem \eqref{problem} exists via probability simplex projection \cite{paper13}. Let $\tilde{p}_{o,1}^M$ denote the ordered sequence of $\tilde{p}_1^M$ such that $\tilde{p}_{o,i} \geq \tilde{p}_{o,j}$ for $j \geq i$. Omitting detailed derivations, the solution is given by:
\begin{align}\label{probability_solution}
p_j^*=&\max(\tilde{p}_j-\lambda^*\Vert \mathbf{x}_j\Vert^p-\delta^*,0)\\
\delta^*=&\frac{P_0-B-C/A+AW/C}{DC/A-A}\\
\lambda^*=&-\frac{1+D\delta^*-W}{A}
\end{align}
where $A\triangleq\sum_{j=1}^{D}\Vert \mathbf{x}_j\Vert^p$, $B\triangleq\sum_{j=1}^{D}p_j\Vert \mathbf{x}_j\Vert^p$, $C\triangleq\sum_{j=1}^{D}\Vert \mathbf{x}_j\Vert^{2p}$, $W\triangleq\sum_{j=1}^{D}p_j$ and $D$ is the integer such that $\tilde{p}_{o,j}>0,j=1,\cdots,D$ and $\tilde{p}_{o,j}=0,j=D+1,\cdots,M$.

%In the next, we briefly describe how to obtain the locations of two-dimensional constellation points. According to the Lloyd criteria, there are still several methods for quantization. A natural idea is to separately quantize the two axis via the one-dimensional method and then, combine them by Kronecker product operation. However, this scheme will introduce more performance loss, since it forces the Voronoi region of every constellation points to be squared, which cannot generate larger minimum Euclidean distance (MED) between constellation points. In addition, it has been proven that the hexagonal distribution maximizes the MED and is both the optimal solution of covering and packing problems . Thus, we herein utilize the polar coordinate-based quantization scheme. The procedures are as follows: 1) Determine the radius version of the continuous input distribution; 2) Perform the quantization along the radius axis with the one-dimensional method; 3) For every shell, quantize the constellation points uniformly.

\vspace{-0.2cm}
\section{simulations}
This section validates the proposed channel capacity bounds and evaluates the performance of the low-complexity constellation shaping scheme. The generalized signal-to-noise ratio (GSNR) is defined as $\text{GSNR(dB)} = 10\log_{10}\frac{E(\Vert \mathbf{x}_j\Vert^p)}{2(\gamma_g^2+\gamma_s^2)}$, where $p < \alpha$.

\begin{figure}[htbp]
\centering
\subfloat[$\alpha=1.2$, $\rho=0.2$]{\includegraphics[width=1.7in]{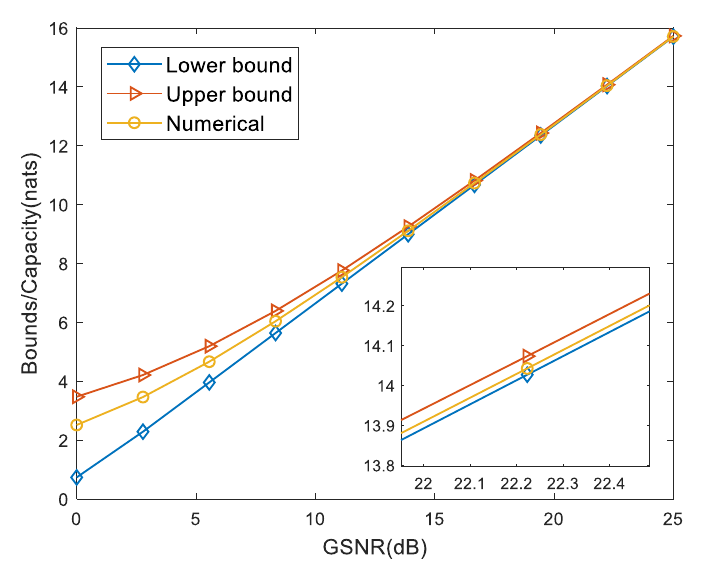}\label{fig_1_a}}
\subfloat[$\alpha=1.8$, $\rho=0.8$]{\includegraphics[width=1.7in]{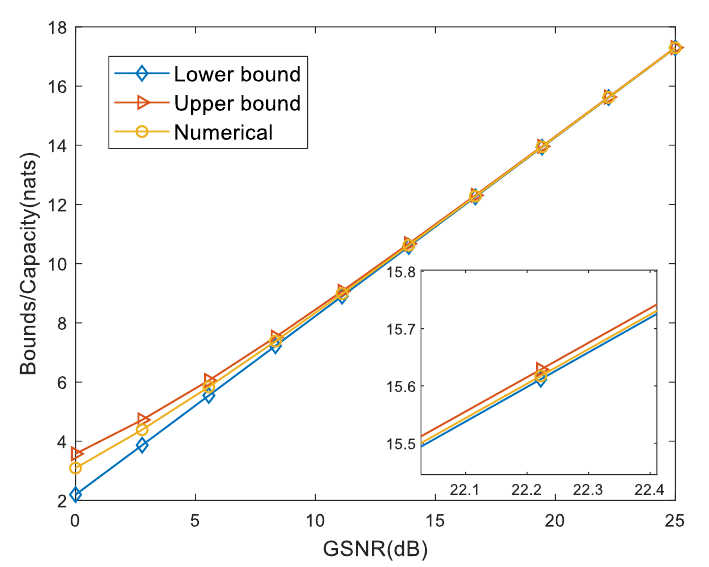}\label{fig_1_b}}
\caption{Comparison of channel capacity and bounds under various $\alpha$ and $\rho$.}
\label{fig_1}
\end{figure}

We first assess the tightness of $C_L$ and $C_U$ by comparing them with the channel capacity computed numerically using the BA algorithm. The parameters are set as $\gamma_g = \gamma_s = 1$, with $(\alpha, \rho)$ values of $(1.2, 0.2)$ and $(1.8, 0.8)$ representing strongly and weakly impulsive scenarios, respectively. We choose $p = 1.1$ to ensure $p < \alpha$. As shown in Fig. \ref{fig_1}, both lower and upper bounds remain tight across different noise configurations. Moreover, $C_L$ and $C_U$ converge to the exact channel capacity as GSNR increases, with the gap diminishing as $\text{GSNR(dB)} \rightarrow +\infty$, consistent with our theoretical analysis.

\vspace{-0.5cm}

\vspace{-0.8cm}
\begin{figure}[htbp]
\centering
\subfloat[$M=16$]{\includegraphics[width=1.7in]{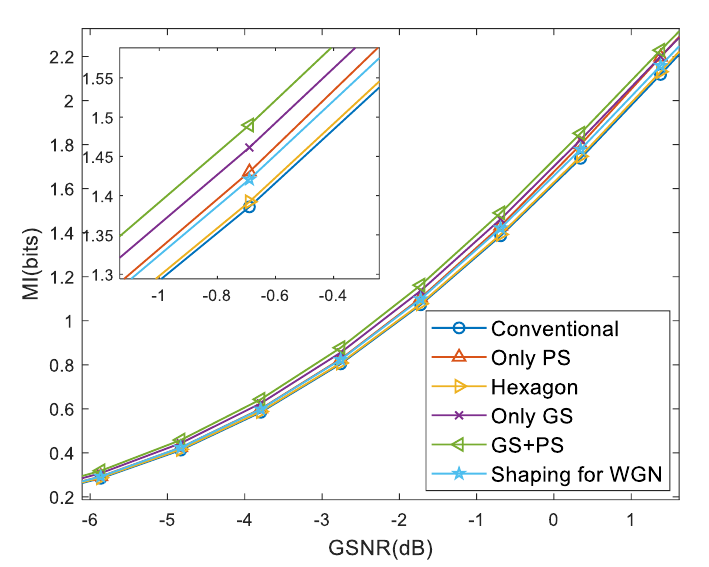}\label{fig_2_a}}
\subfloat[$M=64$]{\includegraphics[width=1.7in]{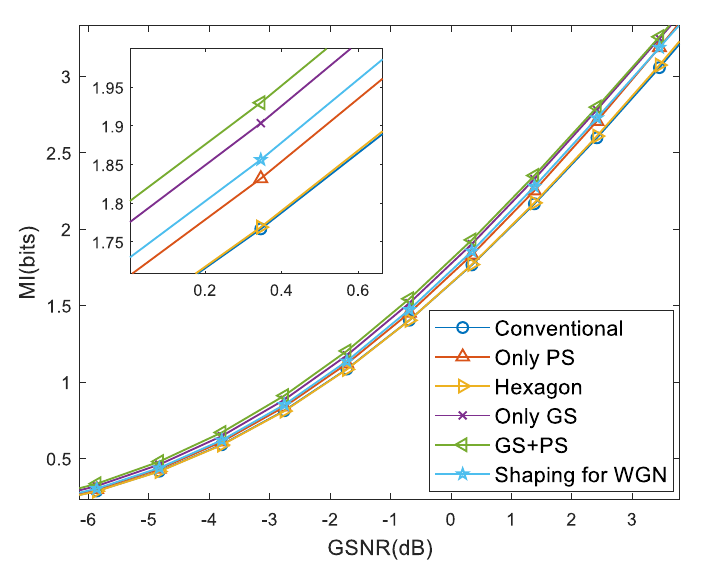}\label{fig_2_b}}
\caption{Mutual information comparison between the low-complexity constellation shaping and baselines under different modulation orders with $(\alpha,\rho)=(1.5,0.5)$.}
\label{fig_2}
\end{figure}

Next, we evaluate the performance of the proposed low-complexity shaping scheme against several baseline methods. The parameters are fixed at $(\alpha, \rho) = (1.5, 0.5)$, with modulation orders $M = 16$ and $M = 64$ considered. The MI comparison is shown in Fig. \ref{fig_2}. `Only PS' and `Only GS' denote that only PS or GS is performed on the conventional constellation. `Hexagon' is the constellation with hexagonally distributed locations and uniform input distribution. `Shaping for WGN' denotes the shaping scheme for the WGN scenario, which applies the MB distribution. Fig. \ref{fig_2} reveals a significant MI loss when the IN component is ignored. Furthermore, the performance gain from GS exceeds that from PS, which aligns with existing results showing that any shaping gain achievable through PS can also be attained via an appropriate GS approach \cite{paper30}. Finally, the MI improvement becomes more pronounced as the modulation order increases, consistent with theoretical expectations.

\vspace{-0.3cm}
\section{conclusions}
This paper investigated channel capacity bounds and their application to constellation optimization under memoryless mixed Gaussian-impulsive noise. We formulated the capacity problem under a $p$-th power constraint and derived closed-form lower and upper bounds. Through asymptotic analysis, an explicit expression for the channel capacity was obtained in the high-power regime. Building on the corresponding optimal input distribution, a low-complexity constellation shaping method was developed. Simulation results demonstrated that the proposed bounds are tight and that the shaped constellations achieve measurable mutual information gain over existing baseline schemes.

\footnotesize
\bibliographystyle{IEEEtran}
\bibliography{ref}

\end{document}